\documentclass[12pt,draftclsnofoot, onecolumn]{IEEEtran}
\usepackage{graphicx,cite,epsfig,amssymb,amsmath,subfigure,url,stfloats,latexsym}
\usepackage{multirow,array}
\usepackage{graphicx,cite,amsmath,amssymb,hhline,booktabs,stfloats,bm}
\usepackage{verbatim}
\usepackage[utf8]{inputenc}
\usepackage{subfigure}
\usepackage{algorithm,algorithmic}

\newtheorem{theorem}{Theorem}

\newtheorem{lemma}{Lemma}
\usepackage{color}
\definecolor{myc1}{rgb}{0,0,1}

\begin{document}

\title{
Energy Efficient Resource Allocation in UAV-Enabled Mobile Edge Computing Networks
}

\author{
\IEEEauthorblockN{Zhaohui Yang, 
Cunhua Pan, 
Kezhi Wang and
Mohammad Shikh-Bahaei
}


\thanks{Z. Yang  and M. Shikh-Bahaei are with the Centre for Telecommunications Research, Department of Informatics, King’s College London, London WC2B 4BG, U.K., (e-mails: yang.zhaohui@kcl.ac.uk; m.sbahaei@kcl.ac.uk).
}

\thanks{C. Pan is with School of Electronic Engineering and Computer Science, Queen Mary University of London, London E1 4NS, U.K.,
 (e-mail: c.pan@qmul.ac.uk).
}
\thanks{K. Wang is with the Department of Computer and Information Sciences, Northumbria University, Newcastle NE2 1XE, U.K., (e-mail:  kezhi.wang@northumbria.ac.uk).
}
}

\maketitle

\begin{abstract}
In this paper, we consider the sum power minimization problem via jointly optimizing user association, power control, computation capacity allocation and location planning in a mobile edge computing (MEC) network with multiple unmanned aerial vehicles (UAVs).
To solve the nonconvex problem, we propose a low-complexity algorithm with solving three subproblems iteratively.
For the user association subproblem, the compressive sensing based algorithm is accordingly is proposed.
For the computation capacity allocation subproblem, the
optimal solution is obtained in closed form.
For the location planning subproblem, the optimal solution is effectively obtained via one-dimensional search method.
To obtain a feasible solution for this iterative algorithm, a fuzzy c-means clustering based algorithm is proposed.
Numerical results show that the proposed algorithm achieves better performance than conventional approaches.

\end{abstract}

\begin{IEEEkeywords}
Unmanned aerial vehicle-enabled communication, mobile edge computing, resource allocation, user association, location optimization.
\end{IEEEkeywords}

\IEEEpeerreviewmaketitle

\newpage

\section{Introduction}

With  high mobility and the explosive growth of data traffic, unmanned aerial vehicles (UAVs) assisted wireless communications have attracted considerable attention \cite{Zeng2017Mag}.
Compared to conventional wireless communications, UAV-enabled wireless communications can provide higher wireless connectivity in areas without infrastructure coverage.
Besides, high throughput can always be achieved in UAV-enabled wireless communications due to the higher probability of line-of-sight (LoS) communication links between user equipments (UEs) and UAVs \cite{7936620,8048502,8247211,8353131}.
Due to the above distinctions, UAVs can be utilized in many applications, such as UAV-enabled relaying \cite{5937283,7959158,8278204,8629002}, UAV-enabled data collection \cite{8119562,8377357,8337901,8329013}, UAV-enabled device-to-device communication networks \cite{7412759,8514812}, UAV-enabled wireless power transfer networks \cite{8365881} and UAV-enabled caching networks \cite{8254370,8614433}.


To fully exploit the design degrees of freedom for UAV-enabled communications, it is crucial to investigate the location and trajectory optimization in UAV-enabled wireless communication networks.
In \cite{6863654}, the altitude of UAV was optimized to provide maximum radio coverage on the ground.
To maximize the number of covered users using the minimum transmit power, an optimal location and altitude placement algorithm  was investigated in \cite{7918510} for UAV-base stations (BSs).
With different quality-of-service (QoS) requirements of users, authors in \cite{8038014} studied the three-dimension UAV-BS placement that maximizes the number of covered users.
Considering the adjustable UAVs' locations, the UAV number minimization was considered in \cite{7762053}.
In \cite{7888557} and \cite{8254657}, the UAV's trajectory was optimized by jointly considering both the communication throughput and the UAV's energy consumption.
Further optimizing user-UAV association, \cite{7875131} investigated the sum power minimization problem of the UAV.
Different from \cite{7918510,8038014,6863654,7762053,7888557,8254657,7875131} with fixed-beamwidth antenna, the beamwidth of the directional antenna was optimized in \cite{He2017CL} with fixed bandwidth allocation to improve the system throughput.
Through jointly optimizing beamwidth and bandwidth, the sum power was further minimized in \cite{8379427}.
Deploying UAVs as users, \cite{mozaffari2018beyond} proposed a novel concept of three-dimensional (3D) cellular networks and developed an optimal 3D cell association scheme \cite{8362271}.

Recently, mobile edge computing (MEC) has been proposed as a promising technology for future communications since it can improve the computation capacity of UEs with computation-hungry  applications, such as, augmented reality (AR) \cite{8016573}.
With MEC, UEs can offload the tasks to the MEC servers that locate at the edge of the network.
Since MEC servers can be deployed near to UEs, network with MEC can provide UEs with low latency and save energy for UEs \cite{7906521}.
There are two operation modes for MEC, i.e., partial and binary computation offloading.
In partial computation offloading, the computation tasks can be divided into two parts, where one part is locally executed and the other part is offloaded to MEC servers \cite{8006982,8254208,7762913,7929399,8240666,8168252,yang2018energy}.
In binary computation offloading, the computation tasks are either locally executed or offloaded to MEC servers \cite{6574874,8334188}.

Due to the mobility of UAVs, the integration of UAV-enabled communication with MEC can further improve the computation performance \cite{7842423,8370877,7932157,7933157,zhou2018ee}.
The UAV-enabled MEC architecture was first proposed in \cite{7842423}, which showed that the computation performance can be improved with UAVs.
Jointly optimizing bit allocation and UAV's trajectory, the authors in \cite{7932157} and \cite{7933157} minimized the total mobile energy consumption while satisfying QoS requirements of the offloaded mobile application.
Considering wireless power transfer, the computation rate maximization problem was studied in \cite{zhou2018ee} for an UAV-enabled MEC wireless powered system, subject to the energy harvesting causal constraint and the UAV's speed constraint.
However, the above works \cite{7932157,7933157,zhou2018ee} all considered only one UAV in the UAV-enabled MEC network even though there always exist multiple UAVs for practical applications.

In this paper, we consider resource allocation in a UAV-enabled MEC network with multiple UAVs.
The objective of this paper is to minimize the sum power consumption of UEs and UAVs.
The main contributions of this paper are summarized as follows:
\begin{enumerate}
  \item   We formulate the sum power minimization problem with latency and coverage constraints via jointly optimizing user association, power control, computation capacity allocation and location planning. To solve the nonconvex sum power minimization problem, an algorithm is proposed by solving three subproblems iteratively. 
  We also provide the complexity analysis of the proposed algorithm.
  \item  For user association problem with $\ell_0$-norm, we apply the compressive sensing based algorithm, where the closed-form solution is given in each iteration.
  \item For computing capacity allocation or location planning,
we first decompose the original problem into multiple small optimization problems.
Then,  the optimal computing capacity allocation is derived   in closed form, while the optimal location planning is obtained via one-dimensional search method.
\end{enumerate}

The rest of the paper is organized as follows.
In Section $\text{\uppercase\expandafter{\romannumeral2}}$, we introduce the system model and sum power minimization formulation.
Two iterative algorithms are addressed in Section $\text{\uppercase\expandafter{\romannumeral 3}}$ and Section $\text{\uppercase\expandafter{\romannumeral 4}}$, respectively.
Some numerical results are shown in Section $\text{\uppercase\expandafter{\romannumeral5}}$
and conclusions are finally drawn in Section $\text{\uppercase\expandafter{\romannumeral6}}$.

\section{System Model}

\begin{figure}[htpb]
\centering
\includegraphics[width=5in]{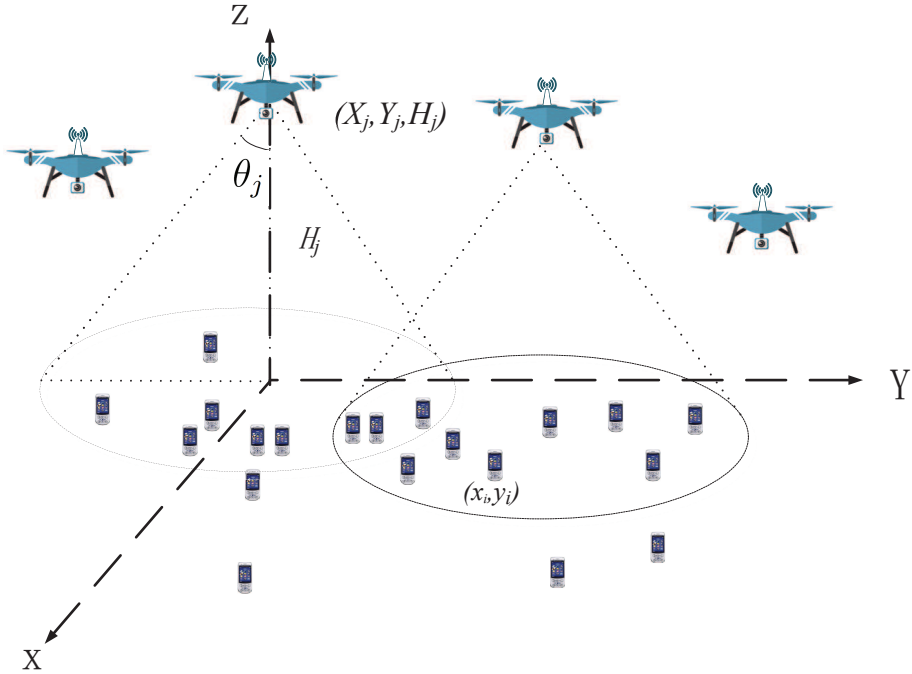}
\caption{UAV-enabled mobile edge computing framework. } \label{uav}
\end{figure}

As shown in Fig. \ref{uav}, we consider a UAV-aided network with $N$ UEs and $M$ UAVs. The sets of UEs and UAVs are denoted by $\mathcal{N}=\{1, 2,..., N\}$ and $\mathcal{M}=\{1, 2,..., M\}$, respectively.  Each UE has a computation task to be executed, which can be offloaded to the UAVs. Define a new set $\mathcal{M'}=\{0, 1, \cdots, M\}$ to represent the possible place in which the tasks can be executed, where $0$ means that UE conducts task itself without offloading. Then, define $a_{ij}$ as the offloading indicator variable from UE $i$ to UAV $j$ satisfying
\begin{equation}\label{w1}
\begin{aligned}
 a_{ij} =\{ 0,1\}, \quad \forall i\in  \mathcal{N}, \forall j\in  \mathcal{M'},
\end{aligned}
\end{equation}
where $a_{ij}=1,\; j \neq0$ denotes that UE $i$ decides to offload the task to UAV $j$, while $a_{ij}=0,\; j \neq0$ indicates that UE $i$ decides not to offload the task to UAV $j$, and $a_{ij}=1,\; j =0$ denotes UE conducts the task itself. One has
\begin{equation}\label{w2}
\begin{aligned}
 \sum_{j=0}^M a_{ij} = 1,\quad i\in \mathcal{N},
\end{aligned}
\end{equation}
which denotes that each task can only be executed at one place.

Similar to \cite{7393804}, we assume that UE $i$ has the computationally intensive task $U_i$ to be executed as follows
\begin{equation}\label{w3}
\begin{aligned}
U_i=(F_i, D_i, T), \quad \forall i\in \mathcal{N},
\end{aligned}
\end{equation}
where $F_i$ describes the total number of the central processing unit (CPU) cycles of $U_i$ to be computed,
$D_i$ denotes the data size transmitting to the cloud if offloading action is decided and $T$ is the latency constraint or QoS requirement by this task. In this paper, we consider that all tasks have the same latency requirement $T$, without loss of generality.
$D_i$ and $F_i$ can be obtained by using the approaches provided in \cite{6253581}.

Then,  the execution time of the task can be calculated as
\begin{equation}\label{w6}
\begin{aligned}
T_{ij}^{\rm C}=\frac{F_i}{f_{ij}}, \quad\forall i\in \mathcal{N}, \forall j\in \mathcal{M'},
\end{aligned}
\end{equation}
where $f_{ij}$ is the computation capacity of UAV $j$ allocated to UE $i$ and $j=0$ means the UE executes the task itself.

If the data is offloaded to the UAV, the time required to offload the data is calculated as
\begin{equation}\label{w7}
\begin{aligned}
T_{ij}^{\rm{Tr}}=\frac{D_i}{r_{ij}}, \quad\forall i\in \mathcal{N},  j\in \mathcal{M},
\end{aligned}
\end{equation}
where $r_{ij}$ is the offloading transmission rate of UE $i$ to UAV $j$.
Then, we can have
\begin{equation}\label{w8}
\begin{aligned}
 a_{ij} \left(\frac{D_i}{r_{ij}} +\frac{F_i}{f_{ij}}\right )\leq T, \quad \forall i\in \mathcal{N}, j\in \mathcal{M},
\end{aligned}
\end{equation}
which means that each task executed in the UAV must meet the latency requirement.
{\color{myc1}{Note that the downloading time from the UAV is low and negligible \cite{7842016}.}}
In (\ref{w8}), we define $a_{ij} \left(\frac{D_i}{r_{ij}} +\frac{F_i}{f_{ij}}\right )=0$ for the case where $a_{ij}=0$ and $f_{ij}=0$.

If this task is executed in UE itself, one has
\begin{equation}\label{12w8}
\begin{aligned}
 a_{ij} \frac{F_i}{f_{ij}}\leq T, \quad \forall i\in \mathcal{N}, j=0.
\end{aligned}
\end{equation}
The computing capacity for the UE $i$ is constrained by
\begin{equation}\label{ww3ww8}
\begin{aligned}
 f_{ij} \leq f_{i,\max}^{\rm{ue}}, \quad \forall i\in \mathcal{N},   j=0.
\end{aligned}
\end{equation}
The power consumption at UE $i$ is given by
\begin{equation}\label{en19}
\begin{aligned}
p^{\rm{ue}}_i=\begin{cases}
 &\!\!\!\!\!\!\sum_{j=1}^{M}  a_{ij} p_{ij},\;\;\;\;
\text{if} \;\;\; \text{offloading},  \\
 & \!\!\!\!\!\! p_i^{\text E} , \qquad\qquad\    \text{ if } \;\;\; \text{local execution}
\end{cases}
\end{aligned}
\end{equation}
where $p_{ij}$ is the transmitting power from UE $i$ to the UAV $j$ and $p_i^{\rm E}$ is the execution power in UE $i$ if UE conducts the task itself, which is given by
\begin{equation}\label{wwwwe8}
\begin{aligned}
p_i^{\rm E} = \kappa_i f_{ij}^{\nu_i}, \quad i\in \mathcal{N},j=0,
\end{aligned}
\end{equation}
where $\kappa_i \geq 0$  and $\nu_i \geq 1$ are positive coefficients specified in the CPU model \cite{7264984}. 
The UE power is constrained by
\begin{equation}\label{wwwww8}
\begin{aligned}
 p^{\rm{ue}}_i \leq P^{\rm{ue}}_{i,max}, \quad i\in \mathcal{N}.
\end{aligned}
\end{equation}

The computing power consumption for UAV $j$ can be given as
\begin{equation}\label{ww8}
\begin{aligned}
p_j^{\rm{uav}}=s_jf_j^{w_j}, \quad \forall j\in \mathcal{M},
\end{aligned}
\end{equation}
where $s_{j}$ and $w_j$ are constants.
In \eqref{ww8}, $f_j$ is the computing capacity provided by UAV $j$ to the associated UEs, which can be given as
\begin{equation}\label{www1w8}
\begin{aligned}
f_j = \sum_{i=1}^{N} a_{ij} f_{ij} ,  \quad \forall  j\in \mathcal{M}.
\end{aligned}
\end{equation}
Due to limited computation capacity,
the computing capacity for the UAV $j$ is constrained by
\begin{equation}\label{w2ww8}
\begin{aligned}
f_j \leq f_{j,\max}^{\rm{uav}}, \quad \forall j\in \mathcal{M}.
\end{aligned}
\end{equation}

Assume that the coordinates of UE $i$ are $(x_i,y_i)$ and the coordinates of UAV $j$ are $(X_j, Y_j,  H_j)$. The horizontal distance between UE $i$ and UAV $j$ is calculated as
\begin{equation}\label{wawew8}
\begin{aligned}
R_{ij}=\sqrt{(X_j-x_i)^2+(Y_j-y_i)^2},  \quad \forall  i\in \mathcal{N}, \forall  j\in \mathcal{M}.
\end{aligned}
\end{equation}

{\color{myc1}{
It is assumed that each UAV is equipped with a directional antenna of adjustable beamwidth.
The azimuth and elevation half-power beamwidths of antenna are equal for UAV $j$, which are both denoted by $2\theta_j\in(0, \pi)$.
According to \cite[Eq.~(2-51)]{constantine2016antenna}, the antenna gain in the direction with azimuth angle $\theta$ and elevation angle $\phi$ can be modelled as
\begin{equation}\label{syseq1}
\begin{aligned}
&G=
  \begin{cases}
\frac{G_0}{\theta^2_j} &\!\! \mbox{if $0\leq \theta\leq \theta_j$ and $0\leq \phi\leq \theta_j$} \\
 g \approx 0 &\!\!  \mbox{otherwise},
  \end{cases}
\end{aligned}
\end{equation}
where $G_0\approx2.2846$,
and  $g$ means the channel gain outside the beamwidth of the antenna.
For simplicity, we set $g=0$.
We consider the case that the UEs are located outdoors, and the channel between each UE and UAV is mainly a LoS path.
The uplink channel gain between UE $i$ and UAV $j$ is
\begin{equation}\label{wwew81}
g_{ij}=\frac{g_0}{H_j^2+R_{ij}^2},
\end{equation}
where
$g_0$ is the channel power gain at the reference distance 1 m, i.e., it is assumed that the communication is neglected via the sidelobes.}}


If UE $i$ wants to offload the task to UAV $j$, it has to be
in the coverage area of UAV $j$, i.e.,
\begin{equation}\label{awawew8}
\begin{aligned}
 R_{ij} \leq H_j \text{tan} \theta_j.
\end{aligned}
\end{equation}


According to \eqref{syseq1} and \eqref{wwew81}, if UE $i$ decides to offload the task to UAV $j$, the data rate is given by
\begin{equation}\label{wwew8}
\begin{aligned}
r_{ij}=B \text{log}_2 \left(1+\frac{\alpha p_{ij}}{\theta_j^2 (H_j^2+R_{ij}^2)}\right), \quad \forall i\in \mathcal{N}, j\in \mathcal{M},
\end{aligned}
\end{equation}
where $B$ is the system bandwidth, $\alpha=g_0G_0/\sigma^2$  and $\sigma^2$ is the noise power.
For UAVs with overlapped coverage area, UAVs are allocated with orthogonal frequency resources, which indicates that there is no interference among UAVs.

{\color{myc1}{
According to constraints (\ref{w8}) and (\ref{12w8}), the latency constraints can be combined as
\begin{equation}\label{Re1eq1}
\sum_{j=1}^M  a_{ij} \left(\frac{D_i}{B \text{log}_2 \left(1+\frac{\alpha p_{ij}}{\theta_j^2 (H_j^2+R_{ij}^2)}\right)} +\frac{F_i}{f_{ij}}\right )
+ \frac{a_{i0} F_i}{f_{i0}}\leq T, \quad \forall i \in \mathcal N.
\end{equation}
According to (\ref{w2}), each UE either conducts the task locally or uploads the task to one unique UAV.
If UE $i$ conducts the task locally, i.e., $a_{i0}=1$ and $a_{ij}=0$, $\forall j \in\mathcal M$, equation \eqref{Re1eq1} becomes
\begin{equation}
a_{i0}\frac{F_i}{f_{i0}}\leq T,
\end{equation}
which is the same as equation (\ref{12w8}).
If UE $i$ uploads the task to one unique UAV $j$, i.e., $a_{ij}=1$, $a_{i0}=0$ and $a_{il}=0$, $l\in\mathcal M\setminus\{j\}$, equation \eqref{Re1eq1} becomes
\begin{equation}
a_{ij} \left(\frac{D_i}{B \text{log}_2 \left(1+\frac{\alpha p_{ij}}{\theta_j^2 (H_j^2+R_{ij}^2)}\right)} +\frac{F_i}{f_{ij}}\right )
+ \frac{a_{i0} F_i}{f_{i0}}\leq T,
\end{equation}
which is the same as equation (\ref{w8}) since $r_{ij}$ in defined in (\ref{wwew8}).}}

In practice, the number of UEs associated with one UAV is limited, i.e.,
\begin{equation}\label{wwew82}
\begin{aligned}
 \sum_{i=1}^N a_{ij} \leq U_j, \quad \forall  j\in \mathcal{M},
\end{aligned}
\end{equation}
where $U_j$ is the maximal allowed number of UEs associated with UAV $j$.

Then, we can formulate the sum power minimization problem as follows:
\begin{subequations}\label{w111}
\begin{align}
\min_{\pmb A, \pmb F, \pmb P, \pmb Z
}\;\;\; & {\color{myc1}{W_1}} \sum_{i=1}^N \sum_{j=1}^ M a_{ij} p_{ij}
+ {\color{myc1}{W_1}} \sum _{i=1} ^N a_{i0} \kappa_i f_{i0}^{\nu_i}
+{\color{myc1}{W_2}}\sum_{j=1}^M \left(s_j \left(\sum_{i=1}^N a_{ij} f_{ij}\right)^{w_j} +{\color{myc1}{Q_j \left\|\sum_{i=1}^N a_{ij}\right\|_0}}\right)
\\ \text{s.t. }  \:\;\;\;  & \sum_{j=0}^M a_{ij} = 1,\quad i\in \mathcal{N}
\\&
 {\color{myc1}{s_j \left(\sum_{i=1}^N a_{ij} f_{ij}\right)^{w_j}
+  Q_j \left\|\sum_{i=1}^N a_{ij}\right\|_0 \leq P_{j,\max}^{\rm{uav}}, \quad \forall j \in \mathcal M}}
\\& \sum_{j=1}^M  a_{ij} \left(\frac{D_i}{B \text{log}_2 \left(1+\frac{\alpha p_{ij}}{\theta_j^2 (H_j^2+R_{ij}^2)}\right)} +\frac{F_i}{f_{ij}}\right )
+ \frac{a_{i0} F_i}{f_{i0}}\leq T, \quad \forall i \in \mathcal N
\\& R_{ij}=\sqrt{(X_j-x_i)^2+(Y_j-y_i)^2},  \quad \forall  j\in \mathcal{N},  j\in \mathcal{M}
\\&a_{ij} R_{ij} \leq H_j \text{tan}\theta_j, \quad \forall  i\in \mathcal{N},   j\in \mathcal{M}
\\&\sum_{j=1}^{M}  a_{ij} p_{ij}+a_{i0}\kappa_i f_{i0}^{\nu_i}  \leq P_{i,\max}^{\rm{ue}}, \quad \forall i \in \mathcal N
\\& \sum_{i=1}^{N} a_{ij} f_{ij}\leq f_{j,\max}^{\rm{uav}}, \quad \forall j \in \mathcal M
\\&  \sum_{i=1}^N a_{ij} \leq U_j, \quad \forall  j\in \mathcal{M}
\\&  a_{ij} =\{ 0,1\},   f_{i0} \leq f_{i,\max}^{\rm{ue}} \quad \forall i\in  \mathcal{N}, j\in  \mathcal{M'}
\\&  f_{ij}\geq 0,p_{ij}\geq 0, H_j^{\min} \leq H  \leq H_j^{\max}, \theta_j^{\min} \leq \theta_j  \leq \theta_j^{\max}, \quad \forall i\in  \mathcal{N},  j\in  \mathcal{M},
\end{align}
\end{subequations}
where $\pmb A=\{a_{ij}\}_{i\in\mathcal N, j \in \mathcal M'}$, $\pmb F=\{f_{ij}\}_{i\in\mathcal N, j \in \mathcal M'}$, $\pmb P=\{p_{ij}\}_{i\in\mathcal N, j \in \mathcal M}$, {\color{myc1}{$\pmb Z=\{X_j, Y_j, H_j, \theta_j\}_{j\in\mathcal M}$}},
{\color{myc1}{$W_1$ and $W_2$ are respectively constant positive weights for UE power and UAV power,
$Q_j$ is the propulsion
power for ensuring the UAV $j$ to remain aloft}},
 $\|\cdot\|_{0}$ is the $\ell_0$-norm,
 and  $P_{j,\max}^{\rm{uav}}>Q_j$ is the maximal battery power of UAV $j$.
$[H_j^{\min},   H_j^{\max}]$ is the feasible region of height $H_j$ determined by obstacle heights and authority regulations, and
$[\theta_j^{\min}, \theta_j^{\max}]$
is the feasible region of half-beamwidth $\theta_j$ determined by practical
antenna beamwidth tuning technique.
{\color{myc1}{The term $Q_j \left\|\sum_{j=1}^N a_{ij}\right\|_0 $ stands for the propulsion power of UAV $j$ if it serves at least one UE.}}

Objective function (\ref{w111}a) is the sum power of UEs and UAVs including transmission power, execution power and propulsion power.
Constraints (\ref{w111}b) represent that the UE either conducts the task locally or uploads the task to one unique UAV.
{\color{myc1}{The maximal power constraint for each UAV is shown in (\ref{w111}c)}}.
{\color{myc1}{Since each UE executes the task itself or uploads the task to one and only one UAV according to (\ref{w111}b), the latency requirements for all UEs can be given in (\ref{w111}d).}}
Constraints (\ref{w111}e) and (\ref{w111}f) state that the offloaded UEs should be in the coverage area of the associated UAVs.
The maximal transmission power constraints for UEs are given in (\ref{w111}g).
The maximal computation capacity and maximal associated number of UEs for UAVs are given in (\ref{w111}h) and (\ref{w111}i), respectively.
{\color{myc1}{There are two major differences with Problem (24) and well-known MEC problems in the literature \cite{8329013,7932157,7933157,zhou2018ee}.
The first difference is that this paper considers the UAV-enabled MEC with multiple UAVs, and the battery energy limit for each UAV is also involved.
The other difference is that Problem (24) optimizes the beamwidth and altitude of all UAVs.}}

\section{Proposed Algorithm}
Due to the nonconvex objective function and discrete constraints, Problem (\ref{w111}) is a nonconvex problem.
It is generally hard to effectively obtain a globally optimal solution for this nonconvex problem.
In the following, a joint optimization algorithm is proposed to obtain a suboptimal solution with an iterative mechanism, where a globally optimal solution is obtained for each subproblem.
Specifically, the user association subproblem is first solved due to the fact that the decision variables for user association are discrete.
Based on the obtained user association, the optimal conditions for the transmission power of UEs are obtained, which is helpful in simplifying the original problem.
According to the optimal conditions for the transmission power of UEs, both computing capacity allocation subproblem and location planning subproblem can be decoupled into multiple small-size problems, which fortunately has closed-form optimal solutions.
The analysis of complexity is also provided.

\subsection{Optimal User Association}


{\color{myc1}{
Problem (\ref{w111}) is hard to be solved due to non-smooth $\ell_0$-norm,  which can be approximately solved via a sequence
of weighted $\ell_1$-norm minimizations in compressive sensing according to \cite{7437385}. 
Taking advantage of this technology,
we approximate the $\ell_0$-norm  in the objective function (\ref{w111}a) as
\begin{equation}\label{PAuser1eq1}
\left\|\sum_{i=1}^N a_{ij}\right\|_0\approx
 \delta_{j}^{(n)} \sum_{i=1}^N a_{ij}+\rho_{j}^{(n)},
\end{equation}
with $\delta_{j}^{(n)}$ and $\rho_{j}^{(n)}$ iteratively updated according to
\begin{equation}\label{PAuser1eq2}
 \delta_{j}^{(n)}=\frac{1}{(\sum_{i=1}^Na_{ij}^{(n)}+\tau)\ln(1+\tau^{-1})},
\end{equation}
and
\begin{equation}\label{PAuser1eq2_1}
\rho_{j}^{(n)}=\frac{(\sum_{i=1}^Na_{ij}^{(n)}+\tau)\ln(1+\tau^{-1}\sum_{i=1}^Na_{ij}^{(n)})-\sum_{i=1}^Na_{ij}^{(n)}}
{(\sum_{i=1}^Na_{ij}^{(n)}+\tau)\ln(1+\tau^{-1})},
\end{equation}
where $a_{ij}^{(n)}$ is value of $a_{ij}$ in the $n$-th iteration, and $\tau$ is a constant regularization factor.

For  (\ref{w111}c), it can be equivalently  transformed to
\begin{equation}\label{PAuser1eq2_2}
 s_j \left(\sum_{i=1}^N a_{ij} f_{ij}\right)^{w_j}
 \leq P_{j,\max}^{\rm{uav}} - Q_j, \quad \forall j \in \mathcal M,
\end{equation}
The reason is that, for each UAV $j$,  \eqref{PAuser1eq2_2} is the same as (\ref{w111}c) if there exists at least one $i$ such that $a_{ij}=1$ and \eqref{PAuser1eq2_2} always holds if $a_{ij}=0$ for all $i$.

Denoting $\mathcal M_i=\left\{j\in\mathcal M\left|\frac{ H_j \text{tan}\theta_j}{ R_{ij}}\geq 1\right.\right\}$, we have $a_{ij}=0$ for all $j\in\mathcal M\setminus \mathcal M_i$ according to (\ref{w111}f).
By using new notation $\mathcal M_i$, constraints (\ref{w111}f) can be omitted.
Using new notation $\mathcal M_i$, approximations (\ref{PAuser1eq1}) and temporarily relaxing the integer constraints, Problem (\ref{w111}) with fixed $(\pmb F, \pmb P, \pmb Z)$ can be rewritten as
\begin{subequations}\label{PAuser2}
\begin{align}
\min_{\pmb A, \pmb f}\;\;\; & W_1\sum_{i=1}^N \sum_{j\in\mathcal M_i} a_{ij} p_{ij}
+W_1\sum _{i=1} ^N a_{i0} \kappa_i f_{i0}^{\nu_i}
+W_2\sum_{j=1}^M s_j f_j^{w_j}
+W_2\sum_{j=1}^M Q_j\left(\delta_{j}^{(n)} \sum_{i=1}^N a_{ij}+\rho_{j}^{(n)}\right)
\\ \text{s.t. } \! \;\;\;  & \sum_{j\in\mathcal M_i} a_{ij} = 1,\quad i\in \mathcal{N},
\\&
 s_j f_j^{w_j}
  \leq P_{j,\max}^{\rm{uav}}-Q_j, \quad \forall j \in \mathcal M
  \\& \sum_{j\in\mathcal M_i}  a_{ij} C_{ij}
+ a_{i0} E_i \leq T, \quad \forall i \in \mathcal N
\\&\sum_{j\in\mathcal M_i}  a_{ij} p_{ij}+a_{i0}\kappa_i f_{i0}^{\nu_i}  \leq P_{i,\max}^{\rm{ue}}, \quad \forall i \in \mathcal N
\\&  \sum_{i=1}^N a_{ij} \leq U_j, \quad \forall  j\in \mathcal{M}
\\&f_j=\sum_{i=1}^N a_{ij} f_{ij}, \quad \forall  j\in \mathcal{M}
\\& f_j \leq f_{j,\max}^{\rm{uav}}, \quad \forall j \in \mathcal M
\\&  0\leq a_{ij} \leq1,   \quad \forall i\in  \mathcal{N}, j\in\mathcal{M'},
\end{align}
\end{subequations}
where $\pmb f=\{f_j\}_{j\in\mathcal M}$, $C_{ij}=\frac{D_i}{B \text{log}_2 \left(1+\frac{\alpha p_{ij}}{\theta_j^2 (H_j^2+R_{ij}^2)}\right)} +\frac{F_i}{f_{ij}}$,
$E_{i}=\frac{ F_i}{f_{i0}}$.
In Problem (\ref{PAuser2}), $f_j=\sum_{i=1}^N a_{ij} f_{ij}$ stands for the computing capacity of UAV $j$.
Note that $\pmb t$ in Problem (\ref{PAuser2}) is an auxiliary vector variable,
which helps us design the Lagrangian dual decomposition
method to get integer solutions.
Obviously, Problem (\ref{PAuser2}) is a convex problem with respect to (w.r.t) ($\pmb A, \pmb t$), which can be effectively solved via the dual method \cite{boyd2004convex}.

\begin{theorem}
For Problem (\ref{PAuser2}), the optimal user association $\pmb A$ and auxiliary vector $\pmb f$ can be respectively expressed as
\begin{equation}\label{PAuser2eq2_2}
a_{ij}^*=\left\{ \begin{array}{ll}
\!\!1, &\text{if}\; j =\arg\min_{j\in\mathcal M_i} h_{ij}\\
\!\!0, &\text{otherwise},
\end{array} \right.
\end{equation}
and
\begin{equation}\label{PAuser2eq2_6_2}
f_j^*=\left.\left(\frac{\mu_j}{W_2w_j s_j}\right)^{\frac1 {w_j-1} }\right|_0^{\bar f_{j,\max}^{\rm{uav}}},
\end{equation}
where
\begin{equation}\label{PAuser2eq2_2_2}
h_{ij}=\left\{ \begin{array}{ll}
\!\!
W_1 p_{ij}
+W_2Q_j \delta_j^{(n)}+\beta_i C_{ij}+\gamma_ip_{ij}
+\lambda_j+\mu_j f_{ij}
, &  \forall i\in \mathcal N, j \in \mathcal M_i\setminus \{0\}\\
\!\!
W_1\kappa_i f_{i0}^{\nu_i}+\beta_i E_i
+\gamma_i \kappa_i f_{i0}^{\nu_i}, &\forall i\in \mathcal N, j=0,
\end{array} \right.
\end{equation}
$\{\beta_i\}_{i\in\mathcal N},\{\gamma_i\}_{i\in\mathcal N},\{\lambda_j\}_{j\in \mathcal M}, \{\mu_j\}_{j\in\mathcal M}$ are Lagrange multipliers associated with constraints (\ref{PAuser2}d)-(\ref{PAuser2}g) respectively,
\begin{equation}\label{appAPAuser2eq2_7}
\bar f_{j,\max}^{\rm{uav}}=\min\left\{\left(\frac{P_{j,\max}^{\rm{uav}}-Q_j}{s_j}\right)^{
\frac{1}{w_j}
}, f_{j,\max}^{\rm{uav}}\right\},
\end{equation}
and $a|_b^c=\min\{\max\{a,b\},c\}$.
If there are multiple minimal points in $\arg\min_{j\in\mathcal M_i} h_{ij}$, we will choose any one of them.
\end{theorem}

\itshape \textbf{Proof:}  \upshape
See Appendix A.
 \hfill $\Box$}}

According to \eqref{PAuser2eq2_2}, each UE $i$ selects UAV $j$ with the smallest coefficient $h_{ij}$.
This is because $h_{ij}$ means the power consumption if UE $i$ uploads data to UAV $j$ according to \eqref{PAuser2eq2}.
Note that the value of $\{\alpha_i\}_{i\in\mathcal N},\{\beta_j\}_{j\in\mathcal M}, \{\gamma_i\}_{i\in\mathcal N},\{\lambda_j\}_{j\in \mathcal M}, \{\mu_j\}_{j\in\mathcal M}$ can be determined by the sub-gradient method \cite{bertsekas2009convex}.
The updating procedure can be given by
\begin{eqnarray}
&&\!\!\!\!\!\!\!\!\!\!\!\!\!\!\!\!\!\!
\beta_i=\left[\beta_i+ \phi\left( \sum_{j\in\mathcal M_i}  a_{ij} C_{ij}
+ a_{i0} E_i - T  \right)\right]^+\label{PAuser2eq2_3}\\
&&\!\!\!\!\!\!\!\!\!\!\!\!\!\!\!\!\!\!
\gamma_i=\left[\gamma_i+ \phi\left( \sum_{j\in\mathcal M_i}  a_{ij} p_{ij}+a_{i0}\kappa_i f_{i0}^{\nu_i}  - P_{i,\max}^{\rm{ue}}  \right)\right]^+\\
&&\!\!\!\!\!\!\!\!\!\!\!\!\!\!\!\!\!\!
\lambda_j=\left[\lambda_j+ \phi\left(\sum_{i=1}^N a_{ij} - U_j \right)\right]^+\\
&&\!\!\!\!\!\!\!\!\!\!\!\!\!\!\!\!\!\!
\mu_j =\mu_j + \psi\left(\sum_{i=1}^N a_{ij} f_{ij} - f_j \right),\label{PAuser2eq2_5}
\end{eqnarray}
where $[x]^+=\max\{x,0\}$, and  $\phi>0$ is a dynamically chosen step-size sequence.
We can adopt the typical self-adaptive scheme of \cite{bertsekas2009convex} to choose the dynamic step-size.
By iteratively optimizing $a_{ij}, f_j$ in (\ref{PAuser2eq2_2})-(\ref{PAuser2eq2_6_2}) and updating $\{\beta_i\}_{i\in\mathcal N},\{\gamma_i\}_{i\in\mathcal N},\{\lambda_j\}_{j\in \mathcal M}, \{\mu_j\}_{j\in\mathcal M}$ according to (\ref{PAuser2eq2_3})-(\ref{PAuser2eq2_5}), the optimal solution of Problem (\ref{PAuser2}) can be obtained via the dual gradient method
 with zero {\color{myc1}{duality gap}}.



The compressive sensing based algorithm for solving Problem (\ref{w111}) with fixed $(\pmb F, \pmb P, \pmb Z)$ is given by Algorithm 1, which is equivalent to a majorization-minimization (MM) algorithm that
can be proved to converge by using the same method in \cite[Appendix~A]{7437385}.

\begin{algorithm}[h]
\caption{ {\color{myc1}{Compressive Sensing Based Algorithm for User Association}}}
\begin{algorithmic}[1]
\STATE Initialize a feasible $\pmb A^{(0)}$ of Problem (\ref{w111}) with fixed $(\pmb F, \pmb P, \pmb Z)$ and the iteration number $n=0$.
Obtain the values of $\delta_{j}^{(0)}$ and $\rho_{j}^{(0)}$ according to (\ref{PAuser1eq2}) and
(\ref{PAuser1eq2_1}), respectively.
\REPEAT
\STATE Initialize Lagrange multipliers $\{\beta_i\}_{i\in\mathcal N},\{\gamma_i\}_{i\in\mathcal N},\{\lambda_j\}_{j\in \mathcal M}, \{\mu_j\}_{j\in\mathcal M}$.
\REPEAT
\STATE Obtain the optimal user association $\pmb A$ and auxiliary vector $\pmb f$ according to (\ref{PAuser2eq2_2})-(\ref{PAuser2eq2_6_2}).
\STATE Update Lagrange multipliers  $\{\beta_i\}_{i\in\mathcal N},\{\gamma_i\}_{i\in\mathcal N},\{\lambda_j\}_{j\in \mathcal M}, \{\mu_j\}_{j\in\mathcal M}$ based on (\ref{PAuser2eq2_3})-(\ref{PAuser2eq2_5}).
\UNTIL the objective function (\ref{PAuser2}a) converges
\STATE Denote $(\pmb A^{(n+1)},\pmb f^{(n+1)})$ as the optimal solution of Problem (\ref{PAuser2}).
\STATE Set $n=n+1$, and update the values of $\delta_{j}^{(n)}$ and $\rho_{j}^{(n)}$ according to (\ref{PAuser1eq2}) and
(\ref{PAuser1eq2_1}), respectively.
\UNTIL the objective function (\ref{w111}a) converges
\end{algorithmic}
\end{algorithm}

\subsection{Optimal Power Control}
To solve Problem (\ref{w111}) with given user association $\pmb A$, we have the following lemma for the optimal power control.

\begin{lemma}
For the optimal solution to Problem (\ref{w111}) with given user association $\pmb A$, constraints (\ref{w111}b) always hold with equality, i.e., the optimal power $p_{ij}^*$ can be expressed by
\begin{equation}\label{PApowereq2}
p_{ij}^*=\frac{1}
{\alpha} \left(2^{ \frac{D_i f_{ij}}{B(Tf_{ij}-F_i)}}-1
\right){\theta_j^2 (H_j^2+{(X_j-x_i)^2+(Y_j-y_i)^2})}, \quad \forall j \in \mathcal M, i \in \mathcal N_j,
\end{equation}
where $\mathcal N_j=\{i\in\mathcal N|a_{ij}=1\}$ denotes the set of users associated with UAV $j$, $j\in\mathcal M'$.
\end{lemma}

\itshape \textbf{Proof:}  \upshape
See Appendix B.
 \hfill $\Box$

{\color{myc1}{
Based on Lemma 1, the optimal power $p_{ij}^*$ is a function of computing capacity $\pmb F$, and 3D location $\pmb Z$.
In the following optimization problem, we substitute the optimal power $p_{ij}^*$ given in (\ref{PApowereq2}) into Problem (\ref{w111}).
As a result, Problem (\ref{w111}) with given user association can be effectively solved by optimizing computation capacity and 3D UAV location.}}
\subsection{Optimal Computing Capacity Allocation}
For Problem (\ref{w111}) with fixed user association $\pmb A$ and 3D location $\pmb Z$, 
the computing capacity allocation problem can be formulated as
\begin{subequations}\label{PAComputingeq1}
\begin{align}
\min_{\pmb F}\;\;\; &W_1\sum_{j=1}^ M \sum_{i\in\mathcal N_j} G_{ij}   \left(2^{ \frac{D_i f_{ij}}{B(Tf_{ij}-F_i)}}-1
\right)+
W_1 \sum _{i\in\mathcal N_0}   \kappa_i f_{i0}^{\nu_i}
+ W_2 \sum_{j=1}^M  s_j \left(\sum_{i\in\mathcal N_j}   f_{ij}\right)^{w_j}
\\ \text{s.t. } \!\;\;\;
& \sum_{i\in \mathcal N_j}  f_{ij}\leq \bar f_{j,\max}^{\rm{uav}}, \quad \forall j \in \mathcal M
\\&  f_{i0,\min}\leq   f_{i0} \leq f_{i0,\max}^{\rm{ue}}, \quad \forall i\in  \mathcal{N}_0
\\&  f_{ij}\geq f_{ij,\min}, \quad \forall j\in  \mathcal{M},  i\in  \mathcal{N}_j,
\end{align}
\end{subequations}
where $G_{ij}\!=\!\frac1{\alpha}{{\theta_j^2 (H_j^2+{(X_j-x_i)^2+(Y_j-y_i)^2})}}$,
$\bar f_{j,\max}^{\rm{uav}}$ is defined in \eqref{appAPAuser2eq2_7},
$f_{i0,\min}\!=\!\frac{F_i}T$, $f_{i0,\max}\!=\!\min\left\{\!\left(\frac{P_{i,\max}^{\rm{ue}}}{\kappa_i}\right)^{\frac{1}{\nu_i}},
f_{i,\max}^{\rm{ue}}\!\right\}$,
and
\begin{equation}\label{compuMinCons}
f_{ij,\min}=
\frac{F_i}
{T-\frac{D_i}{B \text{log}_2 \left(1+\frac{  P_{i,\max}^{\rm{ue}}}{G_{ij}}\right)}}.
\end{equation}
Problem (\ref{PAComputingeq1}) is a convex problem.
To show this, we define function $g(x)=\text e^{\frac{1}{x}}$, $x>0$, and we have
\begin{equation}\label{PAComputingeq1_2}
g''(x) =\frac{1}{x^4} (2x+1)\text e^{\frac{1}{x}}> 0, \quad \forall x>0,
\end{equation}
which shows that $g(x)$ is a convex function.
Since $\frac{D_i f_{ij}}{B(Tf_{ij}-F_i)}=\frac{D_i  }{B T }+\frac{D_i F_i }{BT(Tf_{ij}-F_i)}$ and both the second term and third term of objective function (\ref{PAComputingeq1}a) are convex, the objective function (\ref{PAComputingeq1}a) is convex.
Due to the fact that the objective function (\ref{PAComputingeq1}a) is convex and all constraints are convex, Problem (\ref{PAComputingeq1}) is a convex problem.

Observing that the objective function (\ref{PAComputingeq1}a) monotonically increases with $f_{i0}$ and constraints (\ref{PAComputingeq1}d) are box, the optimal $f_{ij}^*$ to Problem (\ref{PAComputingeq1}) is $f_{i0}^*= f_{i0,\min}$, $\forall   i\in\mathcal N_0$.
To solve $\{f_{ij}\}_{j\in\mathcal M, i \in \mathcal N_j}$, Problem (\ref{PAComputingeq1}) can be decoupled into $M$ subproblems since both the objective function and constraints can be decoupled.
For UAV $j$, the computing capacity allocation problem can be formulated as
\begin{subequations}\label{PAComputingeq2}
\begin{align}
\min_{  \{f_{ij}\}_{i\in\mathcal N_j}}\;\;\; & W_1\sum_{i\in\mathcal N_j} G_{ij}  \left(2^{ \frac{D_i f_{ij}}{B(Tf_{ij}-F_i)}}-1
\right)
+ W_2s_j \left(\sum_{i\in\mathcal N_j}   f_{ij}\right)^{w_j}
\\ \text{s.t. } \:\:\;\;\;
& \sum_{i\in \mathcal N_j}  f_{ij}\leq \bar f_{j,\max}^{\rm{uav}}
\\&  f_{ij}\geq f_{ij,\min}, \quad  i\in  \mathcal{N}_j.
\end{align}
\end{subequations}

\begin{theorem}
If $\sum_{i\in \mathcal N_j}h_{ij}^{-1}\left(-W_2 s_jw_{j} (f_{j,\max}^{\rm{uav}})^{w_j-1}\right)|_{f_{ij,\min}}>\bar f_{j,\max}^{\rm{uav}}$,
the optimal computing capacity allocation of Problem \eqref{PAComputingeq2} is
\begin{equation}\label{PAComputingeq2_5}
 f_{ij} =h_{ij}^{-1}\left(-W_2s_jw_{j} (\bar f_{j,\max}^{\rm{uav}})^{w_j-1} -\tau_j\right), \quad \forall i \in \mathcal N_j,
\end{equation}
where $a|_b=\max\{a,b\}$, $h_{ij}^{-1}(f_{ij})$ is the inverse function of $h_{ij}(f_{ij})$,
\begin{equation}\label{PAComputingeq2_2}
h_{ij}(f_{ij})=-\frac{(\ln 2)W_1G_{ij}D_iF_i}
{B(Tf_{ij}-F_i)^2} 2^{ \frac{D_i f_{ij}}{B(Tf_{ij}-F_i)}},
\end{equation}
and $\tau_j$ is the solution of
\begin{equation}\label{PAComputingeq2_6}
\sum_{i\in \mathcal N_j}h_{ij}^{-1}\left(-W_2 s_jw_{j} (\bar f_{j,\max}^{\rm{uav}})^{w_j-1} -\tau_j\right)|_{f_{ij,\min}}=\bar f_{j,\max}^{\rm{uav}}.
\end{equation}

If $\sum_{i\in \mathcal N_j}h_{ij}^{-1}\left(-W_2 s_jw_{j} (f_{j,\max}^{\rm{uav}})^{w_j-1}\right)|_{f_{ij,\min}}\leq \bar f_{j,\max}^{\rm{uav}}$, the optimal computing capacity allocation of Problem \eqref{PAComputingeq2} is
\begin{equation}\label{PAComputingeq3_3_1}
 f_{ij} =\left.h_{ij}^{-1}\left(-W_2s_jw_{j} \nu_j^{w_j-1} \right)\right|_{f_{ij,\min}}, \quad \forall i \in \mathcal N_j,
\end{equation}
where $\nu_j$ is the solution of
\begin{equation}\label{PAComputingeq3_6}
\sum_{i\in \mathcal N_j}\left.h_{ij}^{-1}\left(-W_2s_jw_{j} \nu_j^{w_j-1} \right)\right|_{f_{ij,\min}}- \nu_j=0.
\end{equation}
\end{theorem}

\itshape \textbf{Proof:}  \upshape
See Appendix C.
 \hfill $\Box$

{\color{myc1}{
\subsection{Optimal Location Planning}
It remains to investigate the location planning with fixed association and computing capacity allocation.
With optimized $(\pmb A, \pmb F)$, Problem (\ref{w111}) is equivalent to
\begin{subequations}\label{PALoceq1}
\begin{align}
\min_{\pmb Z
}\;\;\; & \sum_{j=1}^ M \sum_{i\in\mathcal N_j} L_{ij}{( H_j^2 + {(X_j-x_i)^2+(Y_j-y_i)^2})}\theta_j^2
\\ \text{s.t. } \!\;\;\;&
\sqrt {(X_j-x_i)^2+(Y_j-y_i)^2}\leq H_j  \tan \theta_j, \quad \forall  j\in \mathcal{M},   i\in \mathcal{N}_j
\\&  H_j^{\min} \leq H  \leq H_j^{\max},\theta_j^{\min} \leq \theta_j  \leq \theta_j^{\max},  \quad   \forall j\in  \mathcal{M},
\end{align}
\end{subequations}
where $L_{ij}=\frac{1}
{\alpha} \left(2^{ \frac{D_i f_{ij}}{B(Tf_{ij}-F_i)}}-1
\right)$.
Due to decoupled objective function and constraints, Problem (\ref{PALoceq1}) can be decoupled into $M$ subproblems.
For UAV $j$, the location planing problem can be formulated as
\begin{subequations}\label{PALoceq2}
\begin{align}
\min_{ X_j, Y_j, H_j, \theta_j
}\;\;\; & \sum_{i\in\mathcal N_j} L_{ij}{( H_j^2+{(X_j-x_i)^2+(Y_j-y_i)^2})} \theta_j^2
\\ \text{s.t. } \!\;\;\;\;\:\:&
  \sqrt{(X_j-x_i)^2+(Y_j-y_i)^2}\leq H_j  \tan \theta_j, \quad \forall  i\in \mathcal{N}_j\\&  H_j^{\min} \leq H  \leq H_j^{\max},\theta_j^{\min} \leq \theta_j  \leq \theta_j^{\max}.
\end{align}
\end{subequations}

Before solving nonconvex Problem \ref{PALoceq2}, we provide the following lemma.
\begin{lemma}
With fixed beamwidth $\theta_j$, Problem \eqref{PALoceq2} is a convex problem.
\end{lemma}

\itshape \textbf{Proof:}  \upshape
See Appendix D.
 \hfill $\Box$

Given any $\theta_j$, the 3D location Problem \eqref{PALoceq2} is convex according to Theorem 3,
which can be effectively solved via the popular interior point method \cite{boyd2004convex}.
To obtain the optimal value of $\theta_j$, the one-dimension search method is applied.
The optimal location planning algorithm is given in Algorithm 2, where $\xi$ is the stepsize of the one-dimensional search method.

\begin{algorithm}[h]
\caption{Optimal Location Planning}
\begin{algorithmic}[1]
\FOR{$\theta_j=\theta_j^{\min}:\xi:\theta_j^{\max}$}
\STATE Obtain the optimal $(X_j,Y_j,H_j)$ of Problem \eqref{PALoceq2} with given $\theta_j$.
\ENDFOR
\STATE Obtain the optimal $\theta_j$ with the minimal objective value (\ref{PALoceq2}a).
\end{algorithmic}
\end{algorithm}
}}

%
%

\subsection{Iterative Algorithm and Analysis}

\begin{algorithm}[h]
\caption{\!\!: Iterative Association, Computation and Location Algorithm}
\begin{algorithmic}[1]
\STATE Set the initial solution $( \pmb A^{(0)}, \pmb F^{(0)}, \pmb P^{(0)},\pmb Z^{(0)})$,  the tolerance $\epsilon$, the iteration number $t=0$, and the maximal iteration number $T_{\max}$.
\STATE Compute objective value $V_{\text{obj}}^{(0)}=U(\pmb A^{(0)}, \pmb F^{(0)}, \pmb P^{(0)}, \pmb Z^{(0)})$, where $U(\pmb A, \pmb F, \pmb P, \pmb Z)\!=\!W_1 \sum_{i=1}^N \sum_{j=1}^ M a_{ij} p_{ij}
\!+ \!W_1 \sum _{i=1} ^N a_{i0} \kappa_i f_{i0}^{\nu_i}
\!+\!W_2\sum_{j=1}^M \left(s_j \left(\sum_{i=1}^N a_{ij} f_{ij}\right)^{w_j} \!+\!Q_j \left\|\sum_{i=1}^N a_{ij}\right\|_0\right).$
\REPEAT
\STATE Set $t=t+1$.
\STATE With fixed $( \pmb F^{(t-1)}, \pmb P^{(t-1)},\pmb Z^{(t-1)})$, obtain the optimal $\pmb A^{(t)}$ of Problem (\ref{w111}).
\STATE With fixed $( \pmb A^{(t)}, \pmb Z^{(t-1)})$, obtain the optimal $\pmb F^{(t)}$ of Problem (\ref{PAComputingeq1}).
\STATE With fixed $( \pmb A^{(t)}, \pmb F^{(t)})$, obtain the optimal $\pmb Z^{(t)}$ of Problem (\ref{PALoceq1}).
\STATE With given $( \pmb A^{(t)}, \pmb F^{(t)},\pmb Z^{(t)})$, obtain the optimal $\pmb P^{(t)}$ according to  (\ref{PApowereq2}).
\STATE Compute objective value $V_{\text{obj}}^{(t)}=U(\pmb A^{(t)}, \pmb F^{(t)}, \pmb P^{(t)}, \pmb Z^{(t)})$.
\UNTIL $\left|V_{\text{obj}}^{(t)}-V_{\text{obj}}^{(t-1)}\right|\Big/V_{\text{obj}}^{(t-1)}<\epsilon$ or $t>T_{\max}$.
\end{algorithmic}
\end{algorithm}

The iterative procedure for solving Problem (\ref{w111}) is given in Algorithm 3.
The idea is iteratively optimizing user association, computation capacity and location, while the transmission power of UEs is uniquely determined by the user association, computation capacity and location.

{\color{myc1}{
\begin{theorem}
The iterative Algorithm 3 always converges.
\end{theorem}

\itshape \textbf{Proof:}  \upshape
See Appendix E.
 \hfill $\Box$}}

The complexity of Algorithm 3 in each iteration lies in solving Problem (\ref{w111}) with fixed $(\pmb F, \pmb P, \pmb Z)$, Problem (\ref{PAComputingeq1}) and Problem (\ref{PALoceq1}).

To solve user association Problem (\ref{w111})  with fixed $(\pmb F, \pmb P, \pmb Z)$, the compressive sensing based Algorithm 1 is adopted.
In Algorithm 1, the complexity of optimizing user association $\pmb A$ and auxiliary vector $\pmb f$ is $\mathcal O(MN)$ according to (\ref{PAuser2eq2_2})-(\ref{PAuser2eq2_6_2}), and the complexity of updating Lagrange multipliers  $(\{\beta_i\}_{i\in\mathcal N}, \{\gamma_i\}_{i\in\mathcal N},\{\lambda_j\}_{j\in \mathcal M}, \{\mu_j\}_{j\in\mathcal M})$ is also $\mathcal O(MN)$ according to (\ref{PAuser2eq2_3})-(\ref{PAuser2eq2_5}).
As a result, the total complexity of solving Problem (\ref{w111})  with fixed $(\pmb F, \pmb P, \pmb Z)$ is $\mathcal O(L_1L_2MN)$, where $L_1$ is the number of iterations for outer layer in Algorithm 1 and $L_2$ is the number of iterations via the dual method of solving Problem (\ref{PAuser2}).
%

For Problem (\ref{PAComputingeq1}), it can be decoupled into $M$ subproblems.
To solve each subproblem (\ref{PAComputingeq2}), the complexity is $\mathcal O(N\log_{2}(1/\epsilon_1))\log_2(1/\epsilon_2)$, where $\mathcal O(1/\epsilon_1)$ is the complexity of obtaining the inverse function $h_{ij}^{-1}(\cdot)$, and $\mathcal O(1/\epsilon_2)$ is the complexity of solving (\ref{PAComputingeq2_6}) or (\ref{PAComputingeq3_6}) via the bisection method.
Hence, the complexity of solving Problem (\ref{PAComputingeq1}) is $\mathcal O(MN\log_{2}(1/\epsilon_1)\log_2(1/\epsilon_2))$.

For  Problem (\ref{PALoceq1}), it can be also decomposed into $M$ subproblems.
To solve subproblem \eqref{PALoceq2}, the optimal location planning Algorithm 2 is applied.
Since  Problem \eqref{PALoceq2} with fixed $\theta_j$  is convex and the  number of variables of this convex problem is three, the complexity of solving  Problem \eqref{PALoceq2} with fixed $\theta_j$ is small and can be neglected.
As a result, the complexity of Algorithm 2 is $\mathcal O((\theta_j^{\max}-\theta_j^{\min})/\xi)$ and the complexity of solving Problem (\ref{PALoceq1}) is $\mathcal O(M(\theta_j^{\max}-\theta_j^{\min})/\xi)$.

Consequently, the total complexity of Algorithm 3 is $\mathcal O(L_0L_1L_2MN+L_0M(\theta_j^{\max}-\theta_j^{\min})/\xi+L_0MN\log_{2}(1/\epsilon_1)\log_2(1/\epsilon_2))$, where $L_0$ denotes the number of outer iterations of Algorithm 3.

\subsection{Fuzzy C-Means Clustering Based Algorithm for Initial Solution}
{\color{myc1}{
Since the feasible set of Problem \eqref{w111} is nonconvex due to constraints (\ref{w111}c)-(\ref{w111}h), there is no standard method to even obtain an initial feasible solution of Problem (\ref{w111}).
In the following, a fuzzy c-means (FCM) clustering based algorithm is proposed to obtain a feasible solution of Problem \eqref{w111}.
From Problem \eqref{w111}, it is observed that the latency constraints (\ref{w111}d) are vital to be satisfied.

To meet the latency constraints (\ref{w111}d), all UEs are classified into two classes: the latency constraints can be satisfied or not when the UE conducts the task itself.
Denote $a_{i0}=1$ and $a_{ij}=0$ for all $j\in\mathcal M$, latency constraints reduce to
\begin{equation}\label{fcmINieq1}
f_{i0} \geq \frac{F_i}{T}, \quad \forall i \in \mathcal N,
\end{equation}
and maximal UE transmission power constraints (\ref{w111}g) become
 \begin{equation}\label{fcmINieq2}
\kappa_i f_{i0}^{\nu_i}  \leq P_{i,\max}^{\rm{ue}}, \quad \forall i \in \mathcal N.
\end{equation}
Combining \eqref{fcmINieq1}, \eqref{fcmINieq2} and (\ref{w111}j), we have
\begin{equation}
\frac{F_i}{T}\leq \min\left\{
\left(\frac{P_{i,\max}^{\rm{ue}}}{\kappa_i}\right)^{\frac{1}{\nu_i}}
,f_{i,\max}^{\rm{ue}}
\right\}
\end{equation}
As a result, $\mathcal N_0\triangleq\left\{ i\in\mathcal N \left|
\frac{F_i}{T}\leq \min\left\{
\left(\frac{P_{i,\max}^{\rm{ue}}}{\kappa_i}\right)^{\frac{1}{\nu_i}}
,f_{i,\max}^{\rm{ue}}
\right\}
\right.
\right\}$ is the set of UEs which can execute the tasks itself to meet the latency constraints.

We only need to meet the latency constraints of the set of UEs $\mathcal N_1=\mathcal N \setminus \mathcal N_0$ with the help of UAVs.
To effectively find a feasible solution,  it is recommended to use all $M$ UAVs.
According to latency constraints (\ref{w111}d), low altitude $H_j$ and beamwidth  $\theta_j$ are preferred to establish high channel gains between UAVs and UEs.
With this consideration, all UAVs are deployed with lowest altitude and beamwidth, i.e.,
$H_j=H_j^{\min}$ and $\theta_j=\theta_j^{\min}$ for all $j\in\mathcal M$.

Then, it remains to design the 2D locations $\{X_j,Y_j\}_{j\in\mathcal M}$ of all UAVs.
From the channel gain equation \eqref{wwew81}, it is found that short distance between UAVs and UEs results in high channel gain and low transmission latency.
This motivates us to formulate the FCM clustering problem, which
%
%
%
%
%
%
is proposed to solve the joint user association and 2D location planning problem:
\begin{subequations}\label{FCMeq1}
\begin{align}
\min_{\bar{ \pmb A}, \bar {\pmb Z }
}\;\;\; & \sum_{i \in \mathcal N_1}\sum_{j=1}^M a_{ij}^m  ({(X_j-x_i)^2+(Y_j-y_i)^2}+(H_j^{\min})^2)
\\ \text{s.t. } \!\;\;\;&
  \sum_{j=1}^M a_{ij}  =1, \quad \forall  i\in \mathcal{N}_1.
 \\& a_{ij}\geq 0, \quad \forall i \in \mathcal N_1, j \in \mathcal M,
\end{align}
\end{subequations}
where $\bar{ \pmb A}=\{a_{ij}\}_{i\in\mathcal N_1, j \in \mathcal M}, \bar {\pmb Z }=\{X_j,Y_j\}_{j \in \mathcal M}$, $m> 1$ is a weighting coefficient.
Note that the objective function (\ref{FCMeq1}a) represents the sum squared distance between all UEs  and associated UAVs, which can be regarded as sum transmission power of UEs according to (\ref{PApowereq2}) in Section III-B.
The user association variable $a_{ij}$ is temporally relaxed in Problem (\ref{FCMeq1}).
Based on \cite{bezdek1984fcm}, an iterative algorithm is proposed to solve Problem (\ref{FCMeq1}) via optimizing $\bar{\pmb A}$ with fixed $\bar{\pmb Z}$ and updating $\bar{\pmb Z}$ with given $\bar{\pmb A}$.
Specifically, given location $\bar{\pmb Z}$, the optimal association is
\begin{equation}\label{FCMeq2_1}
a_{ij}=\frac{({(X_j-x_i)^2+(Y_j-y_i)^2}+(H_j^{\min})^2)^{-\frac{1}{m-1}}}
{\sum_{l=1}^M({(X_l-x_i)^2+(Y_l-y_i)^2}+(H_l^{\min})^2)^{-\frac{1}{m-1}}}, \quad \forall i\in \mathcal N_1, j \in \mathcal M,
\end{equation}
which can be obtained through solving the KKT conditions of Problem (\ref{FCMeq1}) with fixed $\bar{\pmb Z}$.
With optimized $\bar{\pmb A}$, the location is updated by
\begin{equation}\label{FCMeq2_2}
X_j=\frac{\sum_{i\in\mathcal N_1} a_{ij}^m x_i}
{\sum_{i\in\mathcal N_1} a_{ij}^m },
Y_j=\frac{\sum_{i\in\mathcal N_1} a_{ij}^m y_i}
{\sum_{i\in\mathcal N_1} a_{ij}^m }, \quad \forall j \in \mathcal M.
\end{equation}

\begin{algorithm}[h]
\caption{\!\!: FCM Clustering Based Algorithm}
\begin{algorithmic}[1]
\STATE Set the initial location $\bar{\pmb Z}^{(0)}$, iteration number $t=1$, $n_j=0$, $\mathcal N_j=\emptyset$, $S_j=0$, $\forall j \in \mathcal M$.
\REPEAT
\STATE With fixed $\bar{ \pmb Z}^{(t-1)} $, obtain the optimal $\bar{\pmb A}^{(t)}$ according to (\ref{FCMeq2_1}).
\STATE With fixed $\bar{\pmb A}^{(t)}$, obtain the optimal $\bar{\pmb Z}^{(t)}$ according to  (\ref{FCMeq2_2}).
\STATE Set $t=t+1$.
\UNTIL the objective function (\ref{FCMeq1}a) converges.
\FOR{$i\in\mathcal N_1$}
\STATE Resort set $\mathcal M$ in descending order according to the value of $a_{ij}^{(t)}$, and denote the resorted set by $\bar {\mathcal M}$.
\FOR {$j\in\bar{\mathcal M}$}
\STATE Compute $f_{ij,\min}$ according to \eqref{compuMinCons} and $\bar f_{j,\max}^{\rm{uav}}$ according to \eqref{appAPAuser2eq2_7}.
\IF {$n_j\leq N_j$, $\sqrt{(X_j^{(t)}-x_i)^2+(Y_j^{(t)}-y_i)^2} \leq H_j^{\min} \tan\theta_j^{\min}$
and $f_{ij,\min}+S_j\leq\bar f_{j,\max}^{\rm{uav}} $}
\STATE $a_{ij}=1$, $a_{il}=0$, $\forall l \in \mathcal M\setminus\{j\}$,
$n_j=n_j+1$,
$\mathcal N_j=\mathcal N_j \cup \{i\}$, $S_j=S_j+f_{ij,\min}$.
\STATE Set the computing capacity as $f_{ij}=f_{ij,\min}$.
\STATE Obtain the power $p_{ij}$ according to  (\ref{PApowereq2}).
\STATE Jump to Step 7.
\ENDIF
\ENDFOR
\ENDFOR
\end{algorithmic}
\end{algorithm}

After obtaining the user association and UAV location by solving Problem (\ref{FCMeq1}),
a feasible computing capacity allocation for Problem \eqref{PAComputingeq2} is given by
\begin{equation}
f_{ij}=f_{ij,\min}, \quad \forall i \in \mathcal N_j.
\end{equation}
and the feasibility condition of Problem \eqref{PAComputingeq2} is
\begin{equation}
\sum_{i\in\mathcal N_j} f_{ij,\min} \leq \bar f_{ij,\max}^{\rm{uav}}.
\end{equation}
Then, the power control can be accordingly determined by Lemma 1 in Section III-B.
As a result, the FCM clustering based algorithm for finding an initial solution is given in Algorithm~4.
In Algorithm~4, $n_j$ and $\mathcal N_j$ respectively denote the number and set of UEs associated with UAV $j$, and $S_j=\sum_{i\in\mathcal N_j} f_{ij,\min} $, which is used to determine whether the computing capacity of UAV $j$ is enough to serve an additional UE.
In Steps 7-15, we associate the UE with the UAV using the maximal value of $a_{ij}$ obtained from solving Problem  (\ref{FCMeq1}) if maximal UE number constraint and computing capacity constraint of this UAV can be satisfied.
}}


\section{Numerical Results}
In this section, numerical results are presented to evaluate the performance of the proposed Algorithm 3 and the benchmark schemes.
We consider an UAV-enabled MEC network with $M=10$ UAVs and $N=100$ UEs.
The bandwidth of the network is $B=1$ MHz.
For each UAV, we set the altitude and beamwidth intervals as
 $H_j^{\min}=10$ m, $H_j^{\max}=50$ m, $\theta_j^{\min}=\pi/6$,
and $\theta_j^{\max}=\pi/3$ rad.
The propulsion power and maximal battery power for each UAV are respectively set as $Q_j=100$ W \cite{7888557} and $P_{j,\max}^{\text{uav}}=110$ W.
For each UE, the maximal transmission power is {\color{myc1}{$P_{i,\max}^{\text{ue}}=17$ dBm}}, and
the maximal computation capacity is $f_{i,\max}^{\text{ue}}=10^8$ cycles/s.
We set the channel power gain at the reference distance $1$ m is $g_0=1.42\times 10^{-4}$, and the noise power $\sigma^2=-169$ dBm/Hz.
For MEC parameters, we set $\mu_1=\cdots=\mu_N=w_1=\cdots=w_M=3$, $\kappa_1=\cdots=\kappa_N=s_1=\cdots=s_M=10^{-28}$ \cite{zhou2018ee}.
We assume equal MEC parameters for all UEs (i.e., $D_i=D$, $F_i=F$, $\forall i \in \mathcal N$), equal height for all UAVs ($H_j=H$, $\forall j \in \mathcal M$), equal maximal number of allowed associated UEs for all UAVs (i.e., $U_j=U$, $\forall j \in \mathcal M$), and equal maximal computation capacity for all UAVs (i.e., $f_{j,\max}^{\text{uav}}=f_{\max}^{\text{uav}}$, $\forall j \in \mathcal M$).
The constant positive coefficients for UE power and UAV power are set as $W_1=10$ and $W_2=1$.
Unless specified otherwise, the system parameters are set as $D=100 $ Kbits, $F=10^7$ CPU cycles, $T=1000$~ms, $U_1=U_2=\cdots=U_M=30$ users, $m=1.2$ in Problem (\ref{FCMeq1}),  and $f_{\max}^{\text{uav}}=10^9$ cycles/s.

{\color{myc1}{
We compare the proposed iterative association, computation and location Algorithm 3 (labelled as `IACL') with the exhaustive search method to obtain a near globally optimal solution of Problem (\ref{w111}) (labelled as `EXH'), which refers to IACL algorithm with 1000 initial starting points,
the successive convex approximation (SCA)-based algorithm (labelled as `SCAEAH') with fixed altitude and height in \cite{7932157}, and the equal computation capacity allocation (ECC) algorithm with optimized user association, power control and location.}}

\begin{figure}[htpb]
\centering
\includegraphics[width=4.0in]{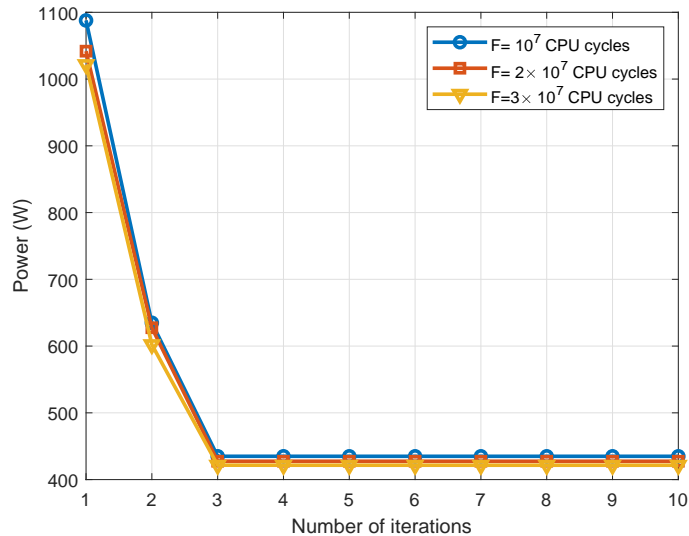}
\caption{Convergence behaviour of the proposed algorithm under different CPU cycles.} \label{fig3}
\end{figure}

Fig. \ref{fig3} illustrates the convergence behaviours for the proposed algorithm under different CPU cycles.
It can be seen that the proposed algorithm converges rapidly, and only three iterations are sufficient to converge, which shows the effectiveness of the proposed algorithm.
The initial solution is high (more than 1000 W), which is due to the fact that the initial solution utilizes all UAVs and the sum propulsion power is high.
After three iterations, the sum power is greatly reduced (nearly 420 W).
This is because the proposed algorithm can efficiently reduce the number of used UAVs and the sum power is thus reduced.

{\color{myc1}{
The sum power of the network versus the maximal latency is depicted in Fig.~\ref{fig5}.
From this figure, it is seen that the sum power decreases with the maximal latency.
This is because large maximal latency allows the UEs and UAVs to transmit with low power.
It is also found that the proposed IACL outperforms the conventional SCAEAH method, since the SCAEAH assumes fixed altitude and beamwidth, while IACL obtains the optimal altitude and beamwidth according to Algorithm 2 in Section III-D.
The proposed IACL also yields better performance than the ECC algorithm with only equal computation capacity allocation, which shows the superiority of the optimization of computation capacity.
Moreover, the EXH algorithm yields the best performance at the sacrifice of high computational complexity.
The gap between the proposed IACL and EXH is small especially for long maximal latency, which indicates that the proposed IACL approaches the near globally optimal solution.

\begin{figure}[htpb]
\centering
\includegraphics[width=4.0in]{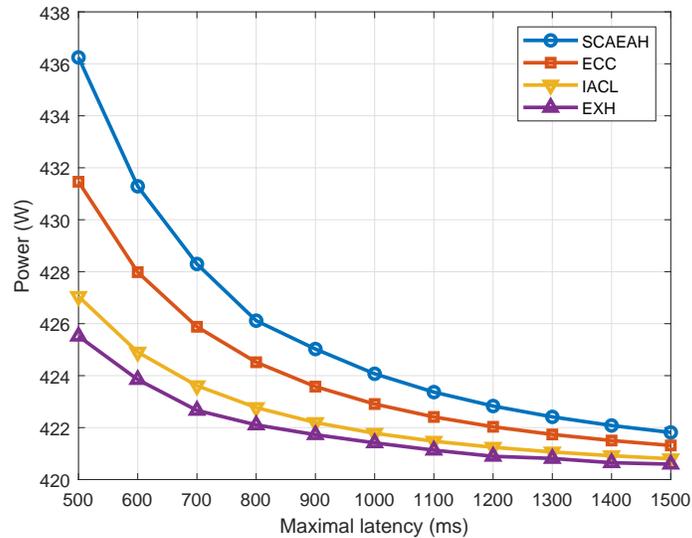}
\caption{Sum power of the network versus the maximal latency $T$.}\label{fig5}
\end{figure}

In Fig.~\ref{fig6}, we illustrate the sum power of the network versus the maximal computation capacity of the UAVs.
It is observed that the sum power decreases with the maximal computation capacity of the UAVs.
This is due to the fact that high computation capacity of the UAVs allows more UEs to offload the traffic to the UAVs, which reduces the power consumption due to the local task computation of the UAVs.
It is also found that the proposed IACL algorithm always outperforms the SCAEAH algorithm, especially for low maximal computation capacity.

\begin{figure}[htpb]
\centering
\includegraphics[width=4.0in]{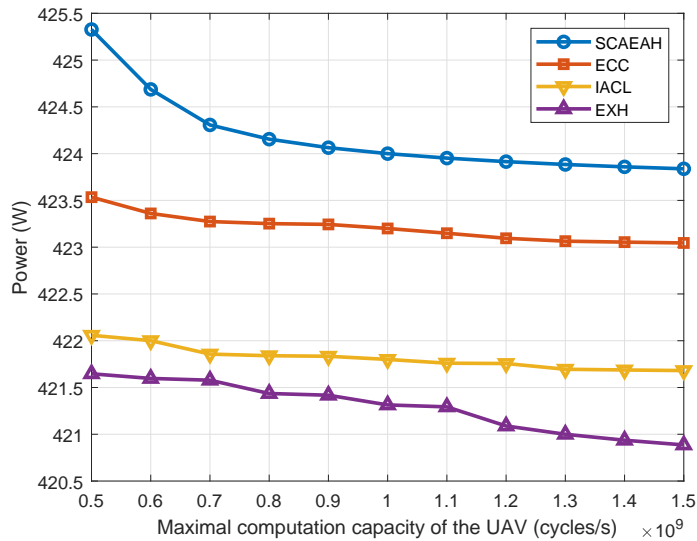}
\caption{Sum power of the network versus the maximal computation capacity of the UAVs $f_{\max}^{\text{uav}}$.} \label{fig6}
\end{figure}}}

\begin{figure}[htpb]
\centering
\includegraphics[width=4.0in]{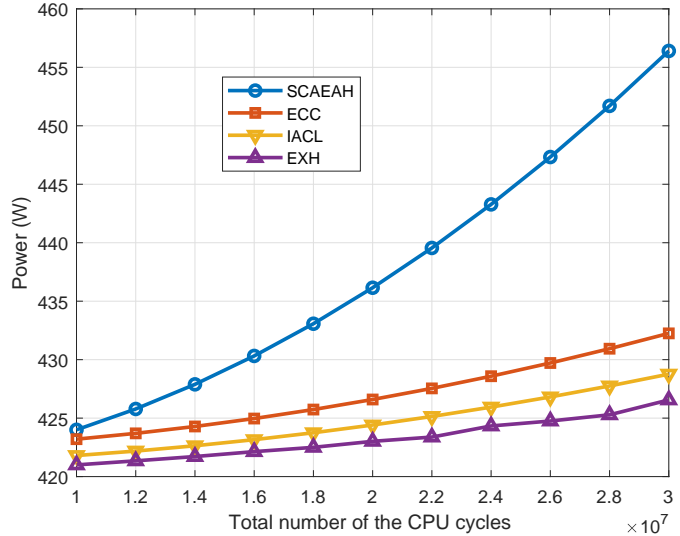}
\caption{Sum power of the network versus total number of the CPU cycles $F$.} \label{fig7}
\end{figure}

The sum power of the network versus total number of the CPU cycles for the tasks that UEs have to be executed is presented in Fig.~\ref{fig7}.
From this figure, we find that the sum power increases with total number of the CPU cycles.
This is because large number of the CPU  cycles requires the UAVs and UEs to allocate high computation capacity to meet the latency constraints, which leads to high power consumption to execute tasks according to  (\ref{w111}a).
It is also shown that the proposed IACL algorithm shows better performance than the SCAEAH algorithm, especially for large CPU cycles.

\begin{figure}[htpb]
\centering
\includegraphics[width=4.0in]{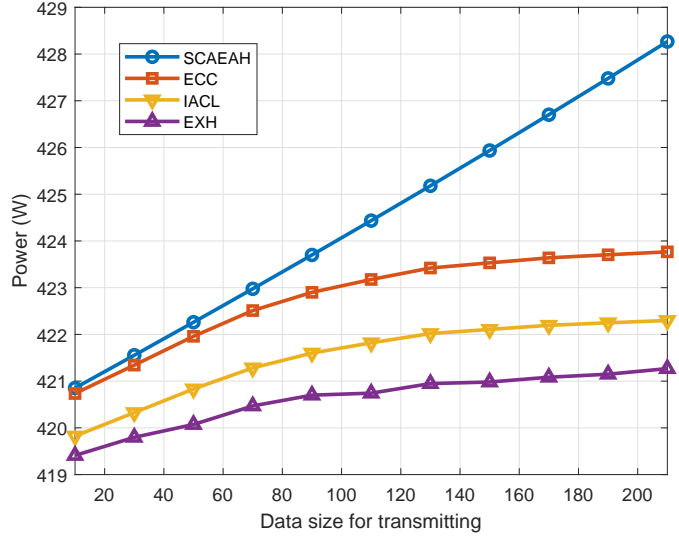}
\caption{Sum power of the network versus the data size $D$.} \label{fig8}
\end{figure}

We show the sum power of the network versus the data size in Fig.~\ref{fig8}.
It is observed that the sum power of the network increases with the data size for all algorithms since more data needs to be computed and more transmission power of the UEs are used to satisfy the latency constraints.
Besides, the grow speed of the sum power versus the data size for the proposed algorithms is slower than that of the SCAEAH algorithm.
Since the proposed IACL algorithm can fully utilize the optimization of latitude and beamwidth, the increased power of UEs for high data rate by IACL is smaller than that by SCAEAH.

\section{Conclusions}
In this paper, we have presented the sum power minimization problem for an UAV-enabled MEC network.
To solve this nonconvex sum power minimization problem, we here proposed an   algorithm through solving three subproblems iteratively.
For user association problem with $\ell_0$-norm, we solved it via the compressive sensing based algorithm.
For computation capacity allocation problem, we decoupled the original problem into multiple problems at small sizes.
The decoupled problems can be proved to be convex ones, and the closed-form solutions were accordingly obtained.
For the location planning problem, the one-dimensional search method was applied to obtain the optimal 3D location and beamwidth.
Numerical results showed that the proposed algorithm achieves better performance than conventional algorithm in terms of sum power consumption, especially for low maximal latency, low maximal computation capacity, high CPU cycles for the tasks and high data rate.
{\color{myc1}{The optimization problem for UAV-enabled MEC network, where UAVs are served as UEs, is left for our future work.}}

\appendices
{\color{myc1}{
\section{Proof of Theorem 1}
\setcounter{equation}{0}
\renewcommand{\theequation}{\thesection.\arabic{equation}}

Denoting $
\pmb \beta=\{\beta_i\}_{j\in\mathcal N}\geq  \pmb 0, \pmb \gamma=\{\gamma_i\}_{i\in\mathcal N}\geq  \pmb 0,\pmb \lambda=\{\lambda_j\}_{j\in \mathcal M}\geq  \pmb 0$ and $\pmb \mu=\{\mu_j\}_{j\in\mathcal M}$  as the Lagrange multiplier vectors associated with constraints (\ref{PAuser2}d)-(\ref{PAuser2}g) respectively, we obtain the dual problem of (\ref{PAuser2}) as
\begin{equation}\label{PAuser2eq1}
\mathop{\max}_{\pmb \beta, \pmb \gamma, \pmb \lambda, \pmb \mu} \quad
D( \pmb \beta, \pmb \gamma, \pmb \lambda, \pmb \mu)=f_{\pmb A}( \pmb \beta, \pmb \gamma, \pmb \lambda, \pmb \mu)+g_{\pmb f}( \pmb \mu),
\end{equation}
where
\begin{equation}\label{PAuser2eq2}
f_{\pmb A}(\pmb \mu)\!=\!\left\{ \begin{array}{ll}
\!\!\!\mathop{\min}\limits_{\pmb{A}}
&\!\!\!\!\!\!
W_1 \sum_{i=1}^N \sum_{j\in\mathcal M_i} a_{ij} p_{ij}
+W_1 \sum _{i=1} ^N \! a_{i0} \kappa_i f_{i0}^{\nu_i}
 \\
&\!\!\!\!\!\!+
W_2\sum_{j=1}^M Q_j\left(\delta_{j}^{(n)} \sum_{i=1}^N a_{ij}+\rho_{j}^{(n)}\right)
+\sum_{i=1}^N \beta_ i \left( \sum_{j\in\mathcal M_i}  a_{ij} C_{ij}
+ a_{i0} E_i - \!T   \right)
 \\
&\!\!\!\!\!\!
+\sum_{i=1}^N \gamma_i \left( \sum_{j\in\mathcal M_i}  a_{ij} p_{ij}+a_{i0}\kappa_i f_{i0}^{\nu_i}  - P_{i,\max}^{\rm{ue}}  \right)
+\sum_{j=1}^M \lambda_j \left(\sum_{i=1}^N a_{ij} - U_j \right)
 \\
&\!\!\!\!\!\!+
\sum_{j=1}^M \mu_j  \sum_{i=1}^N a_{ij} f_{ij}
\\ \text{s.t. } \! \;\;\;  &\!\!\!\!\!\! \sum_{j\in\mathcal M_i} a_{ij} = 1,\quad i\in \mathcal{N}
\\&\!\!\!\!\!\!  0\leq a_{ij} \leq1,   \quad \forall i\in  \mathcal{N}, j\in\mathcal{M'},
\end{array} \right.
\end{equation}
and
\begin{equation}\label{PAuser2eq3}
g_{\pmb f}(\pmb \beta, \pmb \mu)=\left\{ \begin{array}{ll}
\!\!\!\mathop{\min}\limits_{\pmb{f}}
&\!\!
W_2\sum_{j=1}^M s_j f_j^{w_j}
 - \sum_{j=1}^M \mu_j f_j
\\
\textrm{s.t.}
&\!\! s_j f_j^{w_j}
  \leq P_{j,\max}^{\rm{uav}}-Q_j, \quad \forall j \in \mathcal M
\\&\!\! 0 \leq  f_{j}\leq f_{j,\max}^{\rm{uav}}, \quad\forall j  \in {\cal{M}}.
\end{array} \right.
\end{equation}

To minimize the objective function in (\ref{PAuser2eq2}), which is a linear combination of $a_{ij}$,  we should let the association coefficient corresponding to the UAV with the smallest $h_{ij}$  be 1 for any $i$.  Therefore, the solution is thus given as \eqref{PAuser2eq2_2}.

To solve convex Problem (\ref{PAuser2eq3}), we first define $\bar f_{j,\max}^{\rm{uav}}$ in \eqref{appAPAuser2eq2_7}.
Then, the feasible solution of Problem (\ref{PAuser2eq3}) can be simplified as
\begin{equation}\label{appAPAuser2eq2_8}
0\leq  f_{j}\leq\bar f_{j,\max}^{\rm{uav}},\quad\forall j  \in {\cal{M}}.
\end{equation}
For convex Problem (\ref{PAuser2eq3}), we set the first derivative of objective function to zero, i.e.,
\begin{equation}\label{PAuser2eq2_6}
\frac{\partial\left(W_2\sum_{j=1}^M s_j f_j^{w_j}
 - \sum_{j=1}^M \mu_j f_j\right)}
{\partial f_j}=
(W_2+\beta_j)w_j s_j t_j^{w_j-1}-\mu_j=0,
\end{equation}
which yields $f_j=\left(\frac{\mu_j}{W_2w_j s_j}\right)^{\frac1 {w_j-1} }$.
Considering constraints (\ref{appAPAuser2eq2_8}), we can obtain the optimal solution to Problem (\ref{PAuser2eq3}) as \eqref{PAuser2eq2_6_2}.
}}

\section{Proof of Lemma 1}
\setcounter{equation}{0}
\renewcommand{\theequation}{\thesection.\arabic{equation}}

According to constraints (\ref{w111}b), we have
\begin{equation}
p_{ij} \geq \frac{1}
{\alpha}  \left(2^{ \frac{D_i f_{ij}}{B(Tf_{ij}-F_i)}}-1
\right){\theta_j^2 (H_j^2+{(X_j-x_i)^2+(Y_j-y_i)^2})}, \quad \forall j \in \mathcal M, i \in \mathcal N_j
\end{equation}
Since the objective function (\ref{w111}a) increases with $p_{ij}$, the optimal $p_{ij}^*$ can be given by (\ref{PApowereq2}) with any given $(\pmb A, \pmb F, \pmb Z, \pmb \theta)$.
As a result, the optimal $p_{ij}^*$  to Problem (\ref{w111}) with given $\pmb A$ is (\ref{PApowereq2}).

\section{Proof of Theorem 2}
\setcounter{equation}{0}
\renewcommand{\theequation}{\thesection.\arabic{equation}}

Denoting $\tau_j$ as the Lagrange multiplier associated with constraint (\ref{PAComputingeq2}b),
the Lagrangian function of Problem (\ref{PAComputingeq2}) is
\begin{eqnarray}\label{PAComputingeq2kkt}
\mathcal L_1
=W_1\sum_{i\in\mathcal N_j} G_{ij}  \left(2^{ \frac{D_i f_{ij}}{B(Tf_{ij}-F_i)}}-1
\right)
+ W_2 s_j \left(\sum_{i\in\mathcal N_j}   f_{ij}\right)^{w_j}
+\tau_j \left(\sum_{i\in \mathcal N_j}  f_{ij}- \bar f_{j,\max}^{\rm{uav}}\right).
\end{eqnarray}
The Karush-Kuhn-Tucker (KKT) conditions of Problem (\ref{PAComputingeq2}) are:
\begin{subequations}\label{PAcomputingeq2kkt2}
\begin{align}
\!\!\!\!\!\!\!\!\!\!\!\!\!\!\!\!&
\frac{\partial \mathcal L_1}{\partial f_{ij}}=
h_{ij}(f_{ij})
+ W_2s_jw_{j} \left(\sum_{l\in\mathcal N_j}   f_{lj}\right)^{w_j-1} +\tau_j, \quad i \in\mathcal N_j\\
& \tau_j \left(\sum_{i\in \mathcal N_j}  f_{ij}-\bar f_{j,\max}^{\rm{uav}}\right)=0\\
& \sum_{i\in \mathcal N_j}  f_{ij}\leq\bar f_{j,\max}^{\rm{uav}} \\
& \tau_j\geq 0, f_{ij}\geq f_{ij,\min}, \quad  i\in  \mathcal{N}_j,
\end{align}
\end{subequations}
where $h_{ij}(f_{ij})$ is defined in \eqref{PAComputingeq2_2}.
To solve KKT conditions (\ref{PAcomputingeq2kkt2}), we consider the following two cases of $\tau_j$.

1) If $\tau_j>0$, we can obtain
\begin{equation}\label{PAComputingeq2_3}
\sum_{i\in \mathcal N_j}  f_{ij} =\bar f_{j,\max}^{\rm{uav}}
\end{equation}
 according to (\ref{PAcomputingeq2kkt2}b).
From (\ref{PAComputingeq1_2}), function $h_{ij}(f_{ij})$ is a monotonically increasing function.
As a result, substituting (\ref{PAComputingeq2_3}) into (\ref{PAcomputingeq2kkt2}a) and setting $\frac{\partial \mathcal L_2}{\partial f_{ij}}=0$ yield
\begin{equation}\label{PAComputingeq2_5_1}
 f_{ij} =h_{ij}^{-1}\left(-W_2s_jw_{j} (\bar f_{j,\max}^{\rm{uav}})^{w_j-1} -\tau_j\right), \quad \forall i \in \mathcal N_j.
\end{equation}
Considering constraints (\ref{PAcomputingeq2kkt2}d), we further have \eqref{PAComputingeq2_5}.
Combining (\ref{PAComputingeq2_3}) and (\ref{PAComputingeq2_5}), we have \eqref{PAComputingeq2_6}.
Since function $h_{ij}(f_{ij})$ is a monotonically increasing function of $\tau_j$ from (\ref{PAComputingeq1_2}), its inverse function $h_{ij}^{-1}(f_{ij})$ is also a monotonically increasing function, which shows that the left term of function (\ref{PAComputingeq2_6}) is a monotonically decreasing function.
Hence, a unique $\tau_j$ can be obtained via the bisection method.

Having obtained the optimal $\tau_j$ from (\ref{PAComputingeq2_6}), the optimal $f_{ij}$ can be presented in (\ref{PAComputingeq2_5}).
Note that the solution $\tau_j$ to (\ref{PAComputingeq2_6}) should be positive in this case.
To ensure that equation (\ref{PAComputingeq2_6}) has one positive solution, we must have
\begin{equation}\label{PAComputingeq2_6_2}
\sum_{i\in \mathcal N_j}h_{ij}^{-1}\left(-W_2s_jw_{j} (\bar f_{j,\max}^{\rm{uav}})^{w_j-1}\right)|_{f_{ij,\min}}>\bar f_{j,\max}^{\rm{uav}},
\end{equation}
owing to the fact that $h_{ij}^{-1}(f_{ij})$ is a monotonically increasing function.

2) If $\tau_j=0$, we denote
\begin{equation}\label{PAComputingeq3_3}
\sum_{i\in \mathcal N_j}  f_{ij} = \nu_j.
\end{equation}

Substituting (\ref{PAComputingeq3_3}) into (\ref{PAcomputingeq2kkt2}a) yields \eqref{PAComputingeq3_3_1}.
According to (\ref{PAComputingeq3_3}) and (\ref{PAComputingeq3_3_1}), we have
\eqref{PAComputingeq3_6}.
Since the left term of equation  (\ref{PAComputingeq3_6}) is a monotonically decreasing function w.r.t. $\nu_j$, the solution $\nu_j$ to (\ref{PAComputingeq3_6}) can be uniquely obtained via the bisection method.
Based on (\ref{PAcomputingeq2kkt2}c) and (\ref{PAComputingeq3_3}), we have $\nu_j \leq \bar f_{j,\max}^{\rm{uav}}$, which shows that
\begin{equation}
\sum_{i\in \mathcal N_j}h_{ij}^{-1}\left(-W_2s_jw_{j} (\bar f_{j,\max}^{\rm{uav}})^{w_j-1}\right)|_{f_{ij,\min}} -\bar f_{j,\max}^{\rm{uav}} \leq 0.
\end{equation}

\section{Proof of Lemma 2}
\setcounter{equation}{0}
\renewcommand{\theequation}{\thesection.\arabic{equation}}

Define function $\zeta(X_j,Y_j)=\sqrt{(X_j-x_i)^2+(Y_j-y_i)^2}$, and we have
\begin{eqnarray}
\bigtriangledown^2 \zeta(X_j,Y_j)
&&\!\!\!\!\!\!\!\!\!\!
=
\begin{pmatrix}
\frac{\partial ^2 \zeta(X_j,Y_j)}{\partial X_j^2}&\frac{\partial ^2 \zeta(X_j,Y_j)}{\partial X_j\partial Y_j} \\
\frac{\partial ^2 \zeta(X_j,Y_j)}{\partial X_j \partial Y_j}&\frac{\partial ^2 \zeta(X_j,Y_j)}{\partial Y_j^2}
\end{pmatrix}
\nonumber\\
&&\!\!\!\!\!\!\!\!\!\!
=\frac{1}{\left({(X_j-x_i)^2+(Y_j-y_i)^2}\right)^{\frac 3 2}}
\begin{pmatrix}
(X_j-x_i)^2&-(X_j-x_i)(Y_j-y_i) \\
-(X_j-x_i)(Y_j-y_i)&(Y_j-y_i)^2
\end{pmatrix}
\nonumber\\
&&\!\!\!\!\!\!\!\!\!\!
=\frac{1}{\left({(X_j-x_i)^2+(Y_j-y_i)^2}\right)^{\frac 3 2}}
(X_j-x_i,-Y_j+y_i)^T (X_j-x_i, -Y_j+y_i)
\succeq \pmb 0,\nonumber
\end{eqnarray}
which means that function $\zeta(X_j,Y_j)$ is convex and constraints (\ref{PALoceq2}b) are convex.
Since both objective function (\ref{PALoceq2}a) and constraints (\ref{PALoceq1}b) are convex, Problem (\ref{PALoceq2}) is a convex problem.

{\color{myc1}{
\section{Proof of Theorem 3}
\setcounter{equation}{0}
\renewcommand{\theequation}{\thesection.\arabic{equation}}
The proof is established by showing that the sum power (\ref{w111}a) is nondecreasing when sequence ($\pmb x$, $\pmb d$, $\pmb p$) is updated.
According to the IULP algorithm, we have
\begin{eqnarray}\label{Th3convergenceProof}
V_{\text{obj}}^{(t-1)}
&&\!\!\!\!\!\!\!\!
=U(\pmb A^{(t-1)}, \pmb F^{(t-1)}, \pmb P^{(t-1)}, \pmb Z^{(t-1)})
\nonumber\\
&&\!\!\!\!\!\!\!\!
\overset{(\text a)}{\geq}
U(\pmb A^{(t)}, \pmb F^{(t-1)}, \pmb P^{(t-1)}, \pmb Z^{(t-1)})
\nonumber\\
&&\!\!\!\!\!\!\!\!
\overset{(\text b)}{\geq}
U(\pmb A^{(t)}, \pmb F^{(t)}, \pmb P^*(\pmb F^{(t)},\pmb Z^{(t-1)}), \pmb Z^{(t-1)})
\nonumber\\
&&\!\!\!\!\!\!\!\!
\overset{(\text c)}{\geq}
U(\pmb A^{(t)}, \pmb F^{(t)}, \pmb P^*(\pmb F^{(t)},\pmb Z^{(t)}), \pmb Z^{(t)})
\nonumber\\
&&\!\!\!\!\!\!\!\!
=
U(\pmb A^{(t)}, \pmb F^{(t)}, \pmb P^{(t)}, \pmb Z^{(t)})
=V_{\text{obj}}^{(t)},
\end{eqnarray}
where $\pmb P^*(\pmb F, \pmb Z)$ denotes the optimal power function of computing capacity and 3D location as stated in (\ref{PApowereq2}).
Inequality (a) follows from that $\pmb A^{(t)}$ is one suboptimal user association of Problem (\ref{w111}) with fixed computing capacity$\pmb F^{(t-1)}$, power $\pmb P^{(t-1)}$ and location $ \pmb Z^{(t-1)}$.
Inequality (b) is due to the fact that $\pmb F^{(t)}$ is the optimal computing capacity of Problem (\ref{w111}) with fixed user association $\pmb A^{(t)}$ and location $\pmb Z^{(t-1)}$.
Inequality (c) follows from that $\pmb Z^{(t)}$ is the optimal location  of Problem (\ref{w111}) with fixed user association $\pmb A^{(t)}$ and computing capacity $\pmb F^{(t)}$.
Thus, the sum utility is nonincreasing after the update of user association, computing capacity, location and power control.

Furthermore, the sum power (\ref{w111}a) is always positive.
Since the sum power (\ref{w111}a) is nondecreasing in each iteration according to (\ref{Th3convergenceProof}) and the sum power  (\ref{w111}a) is finitely lower-bounded by zero, Algorithm 3 must converge.
}}

\bibliographystyle{ieeetran}
\bibliography{IEEEabrv,MUEC}

\end{document}